\documentclass[journal=ancac3,manuscript=article,layout=twocolumn]{achemso}
\usepackage{verbatim}
\usepackage{sidecap}
\usepackage[mathlines]{lineno}
\usepackage{amsmath,amssymb,amsfonts}
\usepackage[osf]{newpxtext}
\usepackage{newpxmath}
\usepackage[utf8]{inputenx}
\usepackage{graphicx}
\usepackage[dvipsnames]{xcolor}
\usepackage[pdftex]{hyperref}
\usepackage{braket}
\usepackage{xspace}
\usepackage{dblfloatfix}
\usepackage{soul}
\usepackage{chemformula} % Formula subscripts using \ch{}
\usepackage[T1]{fontenc} % Use modern font encodings
\usepackage{siunitx}

\DeclareMathAlphabet{\mathcal}{OMS}{cmsy}{m}{n}

\hypersetup{
	pdfauthor={Philipp John},
	bookmarksnumbered=true,
	pdftitle={ScN/GaN(1$\bar{1}$00): a new platform for the epitaxy of twin-free metal-semiconductor heterostructures},
	colorlinks,
	citecolor=blue,
	linkcolor=blue,
	urlcolor=blue}

\author{P. John}
\email{john@pdi-berlin.de}
\author{A. Trampert}
\author{D.-V. Dinh}
\author{D. Spallek}
\author{J. Lähnemann}
\author{V. Kaganer}
\author{L.\,Geelhaar}
\author{O.\,Brandt}
\author{T.\,Auzelle}
\affiliation[PDI]
{Paul-Drude-Institut für Festkörperelektronik, Leibniz-Institut im Forschungsverbund Berlin e.V., Hausvogteiplatz 5-7, 10117 Berlin}

\title{ScN/GaN(1\={1}00):  a new platform for the epitaxy of twin-free metal-semiconductor heterostructures}


\begin{document}

	%%%%%%%%%%%%%%%%%%%%%%%%%%%%%%%%%%%%%%%%%%%%%%%%%%%%%%%%%%%%%%%%%%%%%
	%% The "tocentry" environment can be used to create an entry for the
	%% graphical table of contents. It is given here as some journals
	%% require that it is printed as part of the abstract page. It will
	%% be automatically moved as appropriate.
	%%%%%%%%%%%%%%%%%%%%%%%%%%%%%%%%%%%%%%%%%%%%%%%%%%%%%%%%%%%%%%%%%%%%%
	%%%%%%%%%%TABLE OF CONTENTS%%%%%%%%%%%%%%%%%%%%%%
	%\begin{tocentry}
	
	%Some journals require a graphical entry for the Table of Contents.
	%This should be laid out ``print ready'' so that the sizing of the
	%text is correct.
	
	%Inside the \texttt{tocentry} environment, the font used is Helvetica
	%8\,pt, as required by \emph{Journal of the American Chemical
		%Society}.
	
	%The surrounding frame is 9\,cm by 3.5\,cm, which is the maximum
	%permitted for  \emph{Journal of the American Chemical Society}
	%graphical table of content entries. The box will not resize if the
	%content is too big: instead it will overflow the edge of the box.
	
	%This box and the associated title will always be printed on a
	%separate page at the end of the document.
	
	%\end{tocentry}
	
	%%%%%%%%%%%%%%%%%%%%%%%%%%%%%%%%%%%%%%%%%%%%%%%%%%%%%%%%%%%%%%%%%%%%%
	%% ABSTRACT
	%% 150 words max.
	%%%%%%%%%%%%%%%%%%%%%%%%%%%%%%%%%%%%%%%%%%%%%%%%%%%%%%%%%%%%%%%%%%%%%
	\begin{abstract}
		\textbf{\small 
			We study the molecular beam epitaxy of rock-salt ScN on the wurtzite GaN(1\={1}00) surface. To this end, ScN is grown on free-standing GaN(1\={1}00) substrates and self-assembled GaN nanowires that exhibit (1\={1}00) sidewalls. On both substrates, ScN crystallizes twin-free thanks to a specific epitaxial relationship, namely ScN(110)[001]$||$GaN(1\={1}00)[0001], providing a congruent, low-symmetry GaN/ScN interface. The 13.1\,\% uniaxial lattice mismatch occurring in this orientation mostly relaxes within the first few monolayers of growth by forming a coincidence site lattice, where 7 GaN planes coincide with 8 ScN planes, leaving the ScN surface nearly free of extended defects.  Overgrowth of the ScN with GaN leads to a kinetic stabilization of the zinc blende phase, that rapidly develops wurtzite inclusions nucleating on \{111\} nanofacets, commonly observed during zinc blende GaN growth. Our ScN/GaN(1\={1}00) platform opens a new route for the epitaxy of twin-free metal-semiconductor heterostructures made of closely lattice-matched GaN, ScN, HfN and ZrN compounds.
		}  
	\end{abstract}
	
	%%%%%%%%%%%%%%%%%%%%%%%%%%%%%%%%%%%%%%%%%%%%%%%%%%%%%%%%%%%%%%%%%%%%%
	%% Introduction
	%%%%%%%%%%%%%%%%%%%%%%%%%%%%%%%%%%%%%%%%%%%%%%%%%%%%%%%%%%%%%%%%%%%%%
	
			%========================================================
	%%Fig.1
	\begin{figure*}[!b]
		\centering
		\includegraphics*[width=\textwidth]{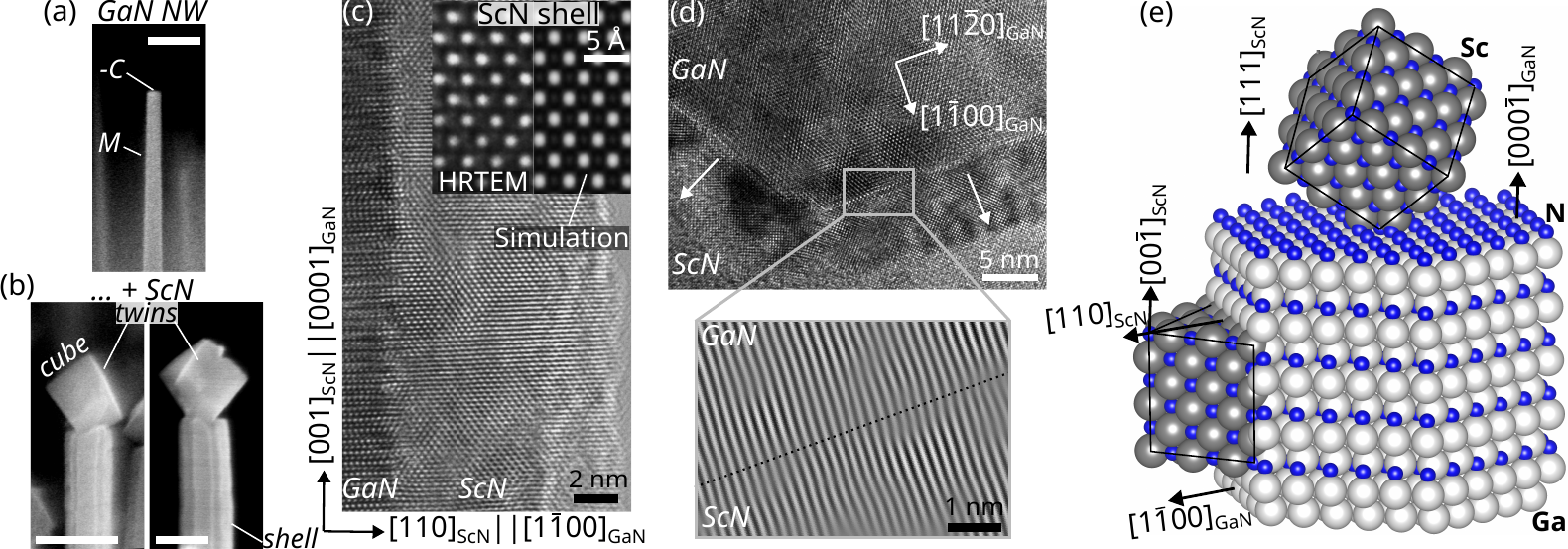}
		\caption{Secondary electron micrographs of GaN NWs (a) before and (b) after overgrowth with a 25\,nm thick ScN shell. The scale bars indicate 200\,nm. (c) HRTEM of the interface between the GaN core and a 10\,nm thick ScN shell with [11\={2}0]$_\text{GaN}$ zone axis. The inset compares a magnified HRTEM image with a simulation using the epitaxial relationship described in the main text. (d) HRTEM along the NW axis, i.e., with [0001]$_\text{GaN}$ zone axis. White arrows within the ScN shell indicate ScN[110] directions. An inverse Fourier-filtered section from the GaN/ScN interface, highlighting GaN(11\={2}0) and ScN(1\={1}0) planes, is shown in the bottom panel. (e) Schematic representation of the local GaN/ScN epitaxial relationships at the NW tip and sidewall.}
		\label{Fig1}
	\end{figure*}
	%========================================================   

	To enhance the functionalities of conventional group III$_A$-nitride (III$_A$-N) devices that are based on the semiconductors AlN, GaN or InN, combinations with transition metal nitrides (TMNs) like NbN, ScN, ZrN or TiN are sought \cite{jena_2019, yan_2018, saha_2018}. These ``new nitrides'' bring in distinctive properties such as superconductivity \cite{yan_2018}, ferroelectricity \cite{fichtner_2019}, catalytic activity \cite{park_2023} or plasmonic resonances in the visible spectrum \cite{dastmalchi_2016,gioti_2020}. This opens up new prospects for enhancing the performance of current devices or creating innovative new ones. To fully realize this potential, successful heterostructuring of the TMNs with III$_A$-N semiconductors will necessitate epitaxial integration with a low density of extended defects. However, while the hexagonal wurtzite phase is the most stable one for III$_A$-nitrides, the cubic rock-salt structure is the common phase for TMNs. Due to the different crystal symmetry, rotational twins are inevitable for the growth of TMNs on the (0001) plane of hexagonal GaN, SiC, or Al$_2$O$_3$  \cite{grundmann_2011}  --- the substrates most commonly used for III$_A$-N heterostructures and devices. Twin domain boundaries locally reduce the crystal symmetry and may thus exhibit dramatically different properties compared to the bulk. In the context of opto-electronic devices, remarkable examples are superconducting twin boundaries in otherwise insulating WO$_{x}$ (with $x \approx 3$), or twin boundaries hosting polarization singularities in otherwise nonpolar perovskite structures \cite{viehland_2014}. Concerning TMNs, preliminary observations suggest that twin boundaries in NbN are not superconducting, which will lower the overall critical temperature of the material \cite{wright_2021}.

	To achieve single-domain TMN layers on hexagonal GaN substrates, we investigate here the epitaxy of ScN on the (1\={1}00) surface (also referred to as \emph{M}-plane surface), since it has a lower symmetry than the conventional (0001) surface (also referred to as \emph{C}-plane surface). (1\={1}00) facets are of practical relevance as they can be realized on large substrates ($\geq2$\,inches) either dislocation-free on the sidewalls of most GaN micro- and nanostructures prepared top-down and bottom-up \cite{largeau_2008,li_2011,koester_2011,pantle_2022}, or with dislocations by heteroepitaxy on $\gamma$-LiAlO$_2$ substrates \cite{waltereit_2000,fernando-saavedra_2023}. Here, ScN is merely used as a model system because it can be grown lattice-matched to GaN \cite{saha_2018,casamento_2019,acharya_2021} and, unlike most transition metals, Sc can be evaporated using an effusion cell. We find that ScN grows free of twins on GaN(1\={1}00) thanks to a specific epitaxial relationship. We obtain this finding by growing twin-free ScN both on free-standing GaN(1\={1}00) substrates and self-assembled GaN nanowires. These two types of substrates facilitate the use of different analytical techniques, yielding consistent results. The uniaxial strain occurring in ScN/GaN(1\={1}00) is relaxed via a coincidence site lattice, making the ScN layers suitable for regrowth of twin-free cubic nitrides.
	
	%%%%%%%%%%%%%%%%%%%%%%%%%%%%%%%%%%%%%%%%%%%%%%%%%%%%%%%%%%%%%%%%%%%%%
	%% Results
	%%%%%%%%%%%%%%%%%%%%%%%%%%%%%%%%%%%%%%%%%%%%%%%%%%%%%%%%%%%%%%%%%%%%%

ScN is first grown by plasma-assisted molecular beam epitaxy (PAMBE) on self-assembled GaN nanowires (NWs) under N-rich conditions to impede the formation of N-vacancies \cite{smith_2001, Dinh2023}. Representative secondary electron micrographs of a GaN NW before and after ScN growth are displayed in Figures\,\ref{Fig1}(a) and \ref{Fig1}(b), respectively. ScN deposited on the (000\={1}) NW top-facet forms a cubic crystallite, whereas ScN grown on the (1\={1}00) NW sidewalls forms a relatively smooth layer. The ScN shell and cube exhibit different crystallographic orientations, which are related to their distinct epitaxial relationships with the underlying GaN facet. As schematized in Figure\,\ref{Fig1}(e), the ScN cube follows the same epitaxial relationship as commonly found for ScN on GaN(0001) substrates, namely ScN(111)[1\={1}0]$||$GaN(000\={1})[11\={2}0] \cite{casamento_2019,acharya_2021,Dinh2023}. Twinning is observed between cubes located on different NWs or even within the same NW (the latter not shown here). In contrast, a very different epitaxial relationship is found for the ScN shell on the GaN NW sidewalls. Figure\,\ref{Fig1}(c) shows a high-resolution transmission electron microscopy (HRTEM) image of the ScN shell taken with the zone axis parallel to the [11\={2}0]$_\text{GaN}$ direction. No twin domains are found after analyzing $\approx$10 NWs. Comparison with the simulation shown in the inset of Figure\,\ref{Fig1}(c) reveals the orientation-relationship ScN(110)[001]$||$GaN(1\={1}00)[0001]. This finding is confirmed by plan-view HRTEM along the [0001]$_\text{GaN}$ zone axis. Figure\,\ref{Fig1}(d) reveals a smooth ScN shell nucleating on well-defined GaN\{1\={1}00\} facets. Since the GaN/ScN orientation-relationship holds on every facet of the NW core, the ScN nucleating on adjacent \{1\={1}00\}$_\text{GaN}$ facets is separated by high-angle boundaries, leading eventually to the formation of vertical grooves as observed in Figure\,\ref{Fig1}(b). The orientation-relationship reported here has already been observed between ScN and AlN as a result of thermal decomposition of (Sc,Al)N thin films \cite{hoglund_2010}, but it is not systematic for all rock-salt compounds grown on wurtzite (1\={1}00) planes (\emph{e.g.}, MgO on ZnO \cite{wu_2012}).

\begin{figure}[!b]
	\includegraphics[width=\columnwidth]{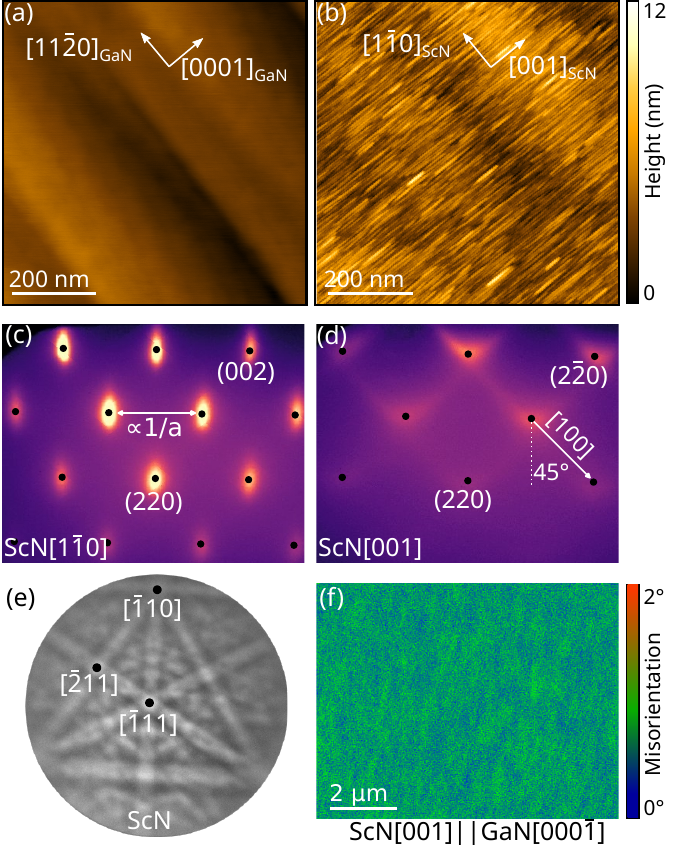}
	\caption{Atomic force topographs of (a) a free-standing GaN(1\={1}00) substrate and (b) a 40\,nm thick ScN layer. (c) and (d) RHEED patterns and superimposed diffraction simulations (black dots) along the ScN azimuths indicated in the bottom left corner. (e) EBSD pseudo-Kikuchi pattern and (f) in-plane orientation map of the ScN layer, where variations in color show the local deviation from the nominal ScN orientation indicated below.}
	\label{Fig2}
\end{figure}

To confirm the epitaxial relationship and the twin-free nature of ScN on a large scale, a $40$\,nm-thick film is grown on a free-standing GaN(1\={1}00) substrate. On such a reference surface, illustrated in the atomic force topograph of Figure\,\ref{Fig2}(a), the ScN orientation is monitored \emph{in situ} by reflection high-energy electron diffraction (RHEED). The patterns taken along the [1\={1}0]$_\text{ScN}$ and [001]$_\text{ScN}$ azimuths (Figures\,\ref{Fig2}(c) and (d), respectively) perfectly match the ones simulated using the epitaxial relationship found for the NW shell and shown by black dots. The presence of $45$\,$^\circ{}$ chevrons in the pattern along the [001]$_\text{ScN}$ azimuth indicates that the layer is terminated by \{100\} facets \cite{Bierwagen2016}. In N-rich conditions, such \{100\} facets are stoichiometric \cite{stampfl_2002,takeuchi_2002} and are typically the ones of lowest energy in rock-salt compounds \cite{marlo_2000}. The nanofaceting of the ScN layer is confirmed by the atomic force topograph shown in Figure\,\ref{Fig2}(b), evidencing an anisotropic surface morphology consisting of periodic stripes, which are oriented along the [001]$_\text{ScN}$ direction. We attribute this anisotropic surface structure to an anisotropy in the adatom diffusion barriers along the ScN [001] and [1\={1}0] directions.

%========================================================
%%Fig.2
%======================================================= %========================================================
%%Fig.3
\begin{figure*}[!b]
	\centering
	\includegraphics[width=\textwidth]{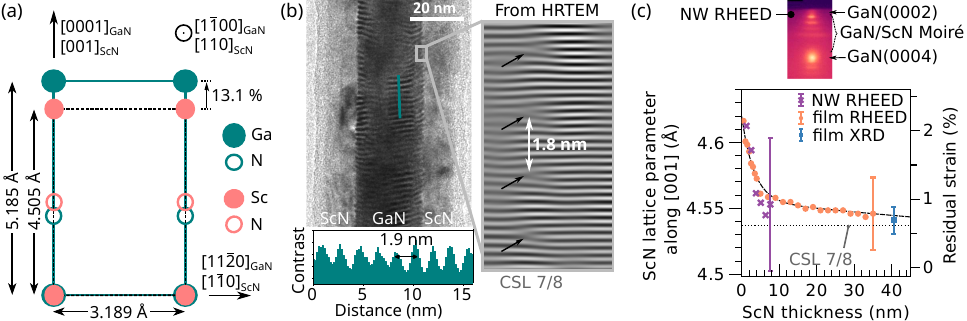}
	\caption{(a) Schematic atomic arrangement of two-dimensional ScN(110) and GaN(1\={1}00) unit meshes. (b) TEM of a GaN/ScN core/shell NW along the [11\={2}0]$_\text{GaN}$ zone axis revealing Moiré patterns. An intensity profile taken along the green line is shown in the bottom panel, and an inverse Fourier-filtered section from a HRTEM image of the interface, highlighting GaN(0002) and ScN(002) planes, is shown in the right panel. Misfit dislocations are indicated by black arrows. (c) ScN lattice parameter measured in films and NWs with RHEED and XRD along the [001]$_\text{ScN}$ direction, as a function of layer thickness. The error bars represent the errors for all data points of the respective measurements. The top panel exemplifies a section of the RHEED pattern during GaN/ScN core/shell NW growth along the [1\={1}00]$_\text{GaN}$ azimuth.}
	\label{Fig3}
\end{figure*}
%========================================================  	

To confirm the absence of small inclusions with a different epitaxial relationship, the ScN film is probed at the \textmu{}m scale with electron backscatter diffraction (EBSD). The pseudo-Kikuchi pattern exemplified in Figure\,\ref{Fig2}(e) is composed of sharp bands, enabling fast and reliable indexing when mapping large areas. Figure\,\ref{Fig2}(f) shows the resulting EBSD orientation map, confirming that the film is single crystalline. The tilt distribution is around $1$\,$^\circ{}$, close to the resolution limit of the technique.

Careful analysis of both the NW shells and the thin films indicates the absence of twin domains. This can be understood by looking at the atomic configuration of the ScN/GaN interface, schematized in Figure\,\ref{Fig3}(a). For the experimentally determined epitaxial relationship, GaN and ScN are lattice-matched along the [11\={2}0]$_\text{GaN}$ direction, but display a $13.1$\,\% mismatch along the [0001]$_\text{GaN}$ direction. No other favorable atomic configuration can be found by rotating or mirroring the ScN lattice in the interfacial plane, and the mirror symmetry of GaN(1\={1}00) matches the one of ScN(110), which are the two necessary conditions to achieve single-domain ScN layers \cite{grundmann_2011}. One competing epitaxial relationship, however, would be that observed for ScN deposited on GaN(0001) (\emph{e.g.}, as seen for MgO on ZnO NWs \cite{wu_2012}). Indeed, for ScN on GaN(1\={1}00), such a configuration would lead to a very low epitaxial strain ($<0.5$\,\%), but the mismatch between the $AB_\text{GaN}$ and $ABC_\text{ScN}$ plane stacking sequence would most likely result in a much higher density of dangling bonds, as explained in the Supporting Information. Therefore, the most favorable interfacial atomic configuration appears here to be the one that minimizes the number of dangling bonds, but with significant epitaxial strain.

The critical thickness for plastic relaxation of the huge uniaxial lattice mismatch ($13.1$\,\%) imposed by the GaN substrate on the ScN layer is less than one monolayer (ML). We thus expect the ScN layer to quickly approach its strain-free lattice parameter taken as $4.505$\,\(\text{\AA}\) \cite{moram_2006}. Consistently, Moiré patterns are observed during the TEM observations of GaN/ScN core/shell NWs [Figure\,\ref{Fig3}(b)], resulting from the superposition of two lattices with different lattice parameters \cite{yokozeki_1974}. The Moiré period along the NW axis ([001]$_\text{ScN}$ direction) is $1.9 \pm 0.1$\,nm, which corresponds to a residual strain of $1.3 \pm 0.7$\,\% for the 12\,nm thick ScN shell. 

Inverse Fourier filtering of an HRTEM image of the ScN/GaN interface is used to highlight the GaN(0002) and ScN(002) planes at the GaN/ScN interface. It evidences a regular array of misfit dislocations with a period of $7~d_{\mathrm{GaN}}^{(0002)} = 8~d_{\mathrm{ScN}}^{(002)} = 1.8$\,nm, as displayed in right panel in Figure\,\ref{Fig3}(b), $d$ indicating the respective inter-planar lattice distance. In contrast, misfit dislocations are absent along the [11\={2}0]$_\text{GaN}$ direction [see bottom panel in Figure \,\ref{Fig1}(d)], where the ScN lattice matches the one of GaN. The dislocations are thus of pure edge (90$^\circ{}$) character and form a $7$/$8$ coincidence site lattice (CSL), for which a residual strain state of $\approx0.6$\,\% is expected. Similar arrays of pure edge dislocations are typically observed in highly-mismatched heterostructures \cite{whaley_1990,trampert_1995, jallipalli_2007,jallipalli_2009}, which do not necessarily lead to a large density of threading dislocations \cite{jallipalli_2009}. Here, no obvious sign for threading dislocations nor stacking faults could be seen within the $\approx10$ NWs examined by TEM.

Next, we closely monitor the kinetics of the strain relaxation using RHEED \cite{coraux_2007}. In the thin film case, the ScN lattice parameter in the [001]$_{\mathrm{ScN}}$ direction is directly retrieved from its reciprocal space lattice measured along the [1\={1}0]$_\mathrm{ScN}$ azimuth [cf. Figure\,\ref{Fig2}(c)] and taking $c_{\mathrm{GaN}} = 5.185$\,\(\text{\AA}\) for the initial GaN substrate \cite{moram_2009}. In the NW case, satellites around the GaN Bragg peaks occur, as displayed in the top panel of Figure\,\ref{Fig3}(c). These satellites are the result of the successive transmission of the electron beam through the NW core and shell, and can be seen as the Fourier transform of the core/shell GaN/ScN Moiré pattern with $p^* = a^* + c^*$, where $p^*$, $a^*$ and $c^*$ are the reciprocal lattice vectors of the Moiré pattern, ScN shell and GaN core, respectively, averaged over $\approx 10^7$\,NWs.

The ScN lattice parameter and residual strain along its [001] axis, extracted from the RHEED analysis of the film and the NWs, are displayed in Figure\,\ref{Fig3}(c) as a function of the ScN thickness. For both layers and NWs, plastic relaxation is observed already at the onset of ScN growth. The strain of the first ScN monolayer is already reduced to $2.0 \pm 0.5$\,\%. Such immediate strain reduction could be explained by the formation of an $8$/$9$ CSL that releases a mismatch of $11.1$\,\%. For further ScN growth of up to $40$\,nm in thickness, strain decreases and tends to saturate at a minimum value of $\approx0.7$\,\%, coinciding with the expected value for a CSL of $7$/$8$ lattice planes for GaN and ScN. This value is also confirmed by X-ray diffractometry (XRD) performed on the $40$\,nm thick ScN film.

A further reduction of the CSL period down to $6$/$7$ lattice planes would overcompensate the mismatch, which does not decrease further the elastic strain energy of the ScN film. Complete strain relaxation therefore implies a break in the CSL periodicity, which is energetically unfavorable and thus explains the apparent saturation of the residual strain at a value of $0.7$\,\% for ScN film thicknesses above $10$\,nm. As a result, the top surface of the $40$\,nm-thick ScN layer grown on GaN(1\={1}00) exhibits a uniaxial strain of $\approx0.7$\,\%, but essentially no extended defects like twin-domain boundaries, stacking faults, or threading dislocations. This is particularly favorable for regrowth of cubic transition metal nitride heterostructures on GaN with high structural perfection.
%========================================================
%%Fig.4
\begin{figure}[H]
	\includegraphics[width=\columnwidth]{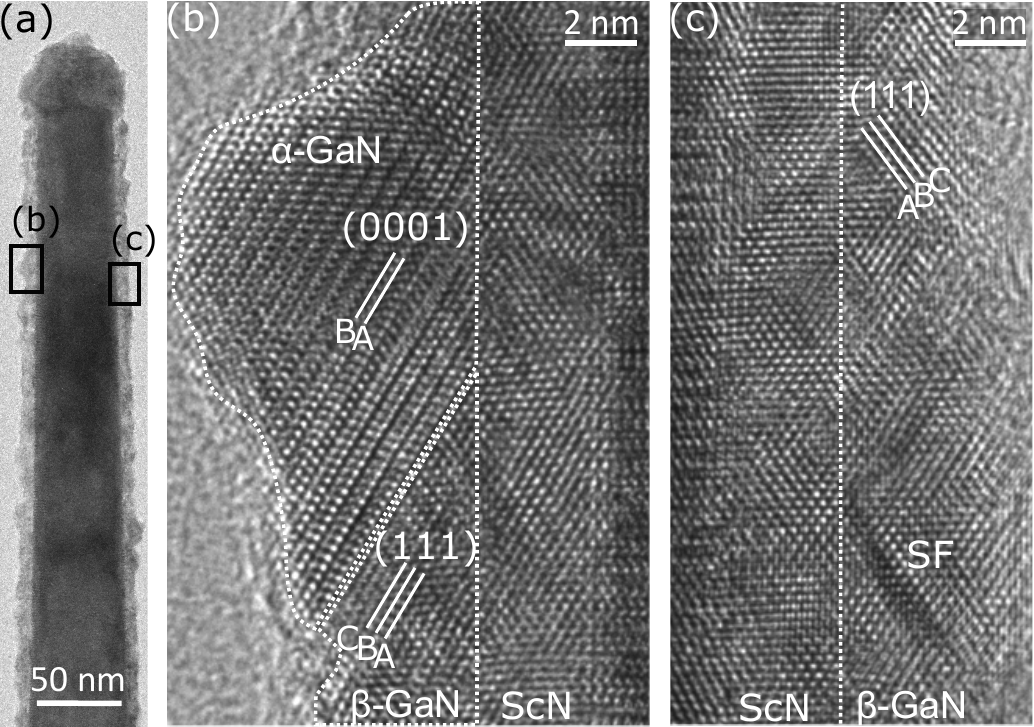}
	\caption{TEM of a GaN/ScN core/shell NW overgrown by GaN, along the [11\={2}0]$_\text{GaN}$ zone axis of the NW core. (a) Overview showing a rough NW shell. Magnified images in (b) and (c) reveal the presence of mixed $\alpha$-GaN/$\beta$-GaN and pure $\beta$-GaN domains, respectively.}
	\label{Fig4}
\end{figure}
%=======================================================	

%========================================================
%%Fig.5

\begin{SCfigure*}
	\centering
	\includegraphics*[width=13cm]{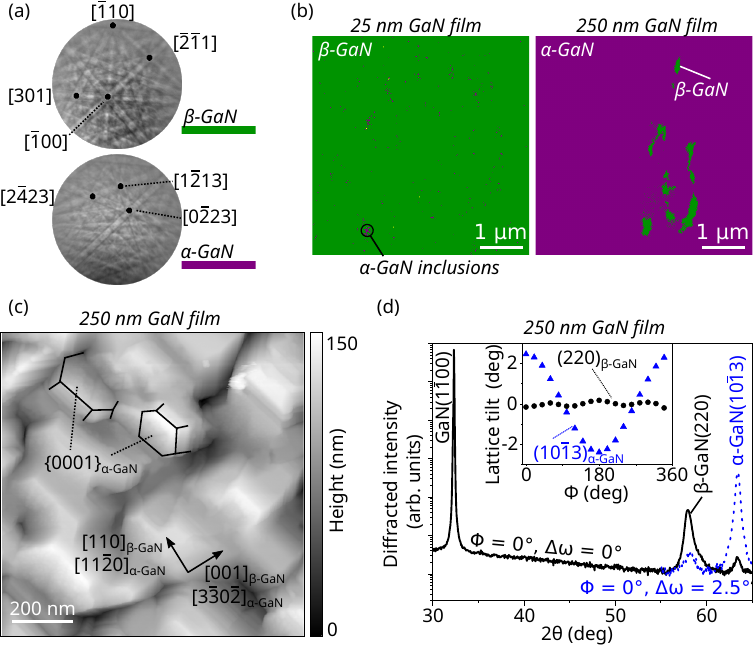}
	\caption{EBSD patterns (a) and phase maps (b) acquired on the $25$ and $250$\,nm thick GaN layers grown on ScN(110)/GaN(1\={1}00) heterostructures. (c)~Atomic force topograph of the $250$\,nm thick GaN layer. (d)~XRD $2\theta/\omega$ scan of the same layer, revealing the presence of $\alpha$-GaN and $\beta$-GaN, tilted relative to each other. The inset shows the lattice tilt of $\alpha$-GaN and $\beta$-GaN as a function of azimuth $\phi$, evidencing $\approx2.5^\circ{}$ of $\alpha$-GaN(10\={1}3) relative to the substrate.} 
	\label{Fig5}
\end{SCfigure*}

%========================================================
	
An interesting question is whether the epitaxial relationship found for ScN grown on GaN(1\={1}00) holds also when GaN is regrown on ScN. To clarify this point, GaN is first deposited on GaN/ScN core/shell NWs and representative TEM micrographs of the resulting GaN/ScN/GaN core/shell structure are shown in Figure\,\ref{Fig4}. The $3$\,nm-thick GaN shell is rougher than the underlying ScN(110) surface and contains both wurtzite ($\alpha$-GaN) and zinc blende ($\beta$-GaN) phases [Figure\,\ref{Fig4}(b)].  The cubic $\beta$-GaN domains share the same orientation as the underlying ScN, namely $\beta$-GaN(110)[001]||ScN(110)[001], resulting in a perfectly commensurate interface [Figure\,\ref{Fig4}(c)]. 
The low energy of the interface is certainly a key factor in explaining the kinetic stabilization of the otherwise thermodynamically unstable zinc blende phase. Surprisingly, the wurtzite crystal phase does not follow the epitaxial relationship found for ScN deposited on GaN(1\={1}00). Instead, $\alpha$-GaN inclusions nucleate on stacking faults (SFs) preferentially formed on \{111\}$_{\beta-\mathrm{GaN}}$ facets, as commonly observed in rough $\beta$-GaN films \cite{trampert_1997}. 
The epitaxial relationship at the wurtzite - zinc blende  interface is then given by $\alpha$-GaN(0001)[11\={2}0]||$\beta$-GaN(111)[1\={1}0]. This translates into the effective orientation of $\alpha$-GaN(1\={1}03)[11\={2}0]||ScN(110)[1\={1}0] with the underlying ScN. The same orientation has been observed for $\alpha$-GaN inclusions occurring during the growth of $\beta$-GaN on MgO(110) \cite{Meyer2022} and GaAs(110) \cite{Saengkaew2015,Daldoul2021}. Here, a $3.3^\circ{}$ tilt between the $\alpha$-GaN[1\={1}03] and ScN[110] directions is expected in the absence of strain.

Continued GaN growth leads to an increase in the size of the wurtzite inclusions, which eventually cover the entire surface. This is evidenced by growing 25 and 250\,nm-thick GaN layers on ScN/GaN(1\={1}00) films. EBSD patterns and phase maps of the two layers are displayed in Figures\,\ref{Fig5}(a) and (b), respectively. The thinner GaN layer exhibits mostly the zinc blende phase, whereas the thicker one shows mostly the wurtzite phase. In spite of the metal-rich growth conditions, the atomic force topograph acquired on the thicker GaN layer and shown in Figure\,\ref{Fig5}(c) reveals a faceted surface, most of them being attributed to wurtzite \{0001\} and \{1\={1}00\} facets, from which we conclude that $\alpha$-GaN\{1\={1}03\} facets are unlikely to be stabilized. Two different in-plane $\alpha$-GaN orientations are found, resulting from the nucleation on both (111) and (11\={1}) facets of the initial $\beta$-GaN layer. 
A symmetric $2\theta$/$\omega$ XRD scan of the 250\,nm GaN film is displayed in Figure\,\ref{Fig5}(d). It evidences the presence of both $\beta$-GaN(220) and $\alpha$-GaN\{1\={1}03\} reflections, the latter being tilted by $2.5^\circ{}$ compared to the $\alpha$-GaN[1\={1}00] substrate direction [see inset of Figure\,\ref{Fig5}(d)]. The reduced tilt compared to the theoretical expectation of $3.3^\circ{}$ is attributed to residual strain in the ScN and top GaN layers. The asymmetry of the $\beta$-GaN(220) peak results from the superposition with the underlying ScN(220) reflection being nearly lattice-matched.
	
Our results indicate that thin III$_A$-nitrides like GaN, AlN and InN can be kinetically stabilized in their metastable $\beta$ phase on ScN(110), thus enabling new metal-semiconductor heterostructures. However, the formation of wurtzite inclusions, a common problem during the growth of cubic III$_A$-nitrides, should be prohibited by using specific growth conditions as previously documented for growth on Si(001) \cite{nishimura_2001}, GaAs(001) \cite{yang_1996}, $3$\emph{C}-SiC \cite{zscherp_2022}. For instance, stabilizing a smooth $\beta$-GaN(110) surface using surfactants \cite{mula_2000} could be decisive to prevent the formation of \{111\}$_{\beta-\mathrm{GaN}}$ facets where $\alpha$-GaN inclusions nucleate. On the other hand, new twin-free TMN heterostructures containing \emph{e.g.} ScN, HfN and ZrN are suitable to be grown on our ScN/GaN(1\={1}00) platform, since TMNs are thermodynamically stable in the cubic rock-salt phase.

%%%%%%%%%%%%%%%%%%%%%%%%%%%%%%%%%%%%%%%%%%%%%%%%%%%%%%%%%%%%%%%%%%%%%
%% conclusions
%%%%%%%%%%%%%%%%%%%%%%%%%%%%%%%%%%%%%%%%%%%%%%%%%%%%%%%%%%%%%%%%%%%%%
	In conclusion, we have studied the growth of GaN/ScN heterostructures on the low-symmetry GaN(1\={1}00) surface, both on NW sidewalls and on free-standing substrates. Twin-free ScN(110) layers are obtained using this approach due to a specific epitaxial relationship for which the interface symmetries match the one of bulk ScN(110). The uniaxial epitaxial strain imposed by the substrate along the GaN[0001] direction plastically relaxes within the first MLs, leaving a residual strain of $\approx0.7$\,\% for ScN layers up to a thickness of $40$\,nm. Subsequent deposition of GaN on ScN/GaN(1\={1}00) results in a kinetic stabilization of $\beta$-GaN that rapidly transforms into a twinned $\alpha$-GaN layer nucleating on \{111\}$_{\beta-\mathrm{GaN}}$ facets. As a result, the epitaxial relationship for $\alpha$-GaN on ScN drastically differs from the one of ScN on $\alpha$-GaN. 
	Importantly, our ScN/GaN(1\={1}00) platform provides a new route for the epitaxy of twin-free heterostructures combining semiconductors and metals like $\beta$-GaN, ScN, HfN and ZrN, which are all nearly lattice-matched \cite{saha_2018}. 
	
%%%%%%%%%%%%%%%%%%%%%%%%%%%%%%%%%%%%%%%%%%%%%%%%%%%%%%%%%%%%%%%%%%%%%
\subsection{Methods}
%%%%%%%%%%%%%%%%%%%%%%%%%%%%%%%%%%%%%%%%%%%%%%%%%%%%%%%%%%%%%%%%%%%%%	

	ScN shells between $3$ and $100$\,nm thickness are grown in N-rich conditions by PAMBE on GaN NWs at $840$\,$^\circ{}$C and with a substrate rotation speed of $2 - 3$\,rpm. Active N is provided by a plasma cell with an atomic flux of $1.5\times10^{15}$\,s$^{-1}$cm$^{-2}$, and Sc by a high-temperature effusion cell with an atomic flux of $1 - 2\times10^{14}$\,s$^{-1}$cm$^{-2}$. The N and Sc fluxes are calibrated by determining the growth rate of GaN and ScN layers grown in N-limited and Sc-limited regimes, respectively. The temperature is calibrated as described in Ref.~\citenum{John2023}. GaN NWs are grown via self-assembly on sputtered TiN films decorated with Si seeds \cite{auzelle_2023}. This approach allows fabricating ensembles with a NW density below $10^9$\,cm$^{-2}$, which is useful to prevent shadowing between neighboring NWs during the shell growth. 
	
	ScN layers with a thickness between $25$ and $40$\,nm are grown on free-standing 5$\times$10\,mm$^2$ GaN(1\={1}00) substrates at 740\,$^\circ{}$C with similar atomic fluxes as used for the NWs. A 100\,nm thick GaN layer is deposited prior to the ScN to create smooth atomic steps. 
	
	GaN overgrowth on ScN is done at 740\,$^\circ{}$C, on NW shells with a III/V ratio of 0.4 and on ScN films in slightly Ga-rich conditions. 
	
	The orientation and lattice parameter of the growing films are monitored \emph{in situ} by RHEED with an acceleration voltage of $20$\,kV  and \emph{ex situ} by XRD using a PANalytical X’Pert PRO MRD diffractometer. EBSD is carried out in a Zeiss Ultra-55 scanning electron microscope operated at $15$\,kV and equipped with an EDAX Hikari Super EBSD camera to identify different crystal phases and orientations. Last, the microstructure of dispersed NWs and of plan-view specimens (preparation described in Ref.\,\citenum{Luna2014}) is  characterized by TEM in a Jeol 2100F field emission microscope operated at 200\,kV and equipped with a Gatan Ultra Scan 4000 charge coupled device. 
	% A specimen for observing the NW cross-section was prepared by standard mechanical grinding and dimpling methods, where final thinning was achieved using a 2\,kV Ar ion beam in a Gatan precision ion polishing system at an incident angle of 4\,$^\circ{}$. 
	Diffraction patterns and HRTEM images are simulated using the JEMS software package, the latter by applying a multi-slice approach.

	\begin{acknowledgement}
		The authors thank Anne-Kathrin Bluhm for SEM measurement, Doreen Steffen for TEM sample preparation and Carsten Stemmler for MBE maintanance. The authors are grateful to Michael Hanke for a critical reading of the manuscript. Funding from Deutsche Forschungsgemeinschaft and Agence Nationale de la Recherche through the project Nanoflex (ANR-21-CE09-0044) is gratefully acknowledged.
		
	\end{acknowledgement}
	
	%%%%%%%%%%%%%%%%%%%%%%%%%%%%%%%%%%%%%%%%%%%%%%%%%%%%%%%%%%%%%%%%%%%%%
	%% The same is true for Supporting Information, which should use the
	%% suppinfo environment.
	%%%%%%%%%%%%%%%%%%%%%%%%%%%%%%%%%%%%%%%%%%%%%%%%%%%%%%%%%%%%%%%%%%%%%
	% \begin{suppinfo}
		% 
		% A listing of the contents of each file supplied as Supporting Information
		% should be included. For instructions on what should be included in the
		% Supporting Information as well as how to prepare this material for
		% publications, refer to the journal's Instructions for Authors.
		% 
		% The following files are available free of charge.
		% \begin{itemize}
			%   \item Filename: brief description
			%   \item Filename: brief description
			% \end{itemize}
		% 
		% \end{suppinfo}
	
	%%%%%%%%%%%%%%%%%%%%%%%%%%%%%%%%%%%%%%%%%%%%%%%%%%%%%%%%%%%%%%%%%%%%%
	%% The appropriate \bibliography command should be placed here.
	%% Notice that the class file automatically sets \bibliographystyle
	%% and also names the section correctly.
	%%%%%%%%%%%%%%%%%%%%%%%%%%%%%%%%%%%%%%%%%%%%%%%%%%%%%%%%%%%%%%%%%%%%%
	\bibliography{ScN_twin-free_paper}
	
\end{document}

\begin{figure*}
	\includegraphics[width=0.75\textwidth]{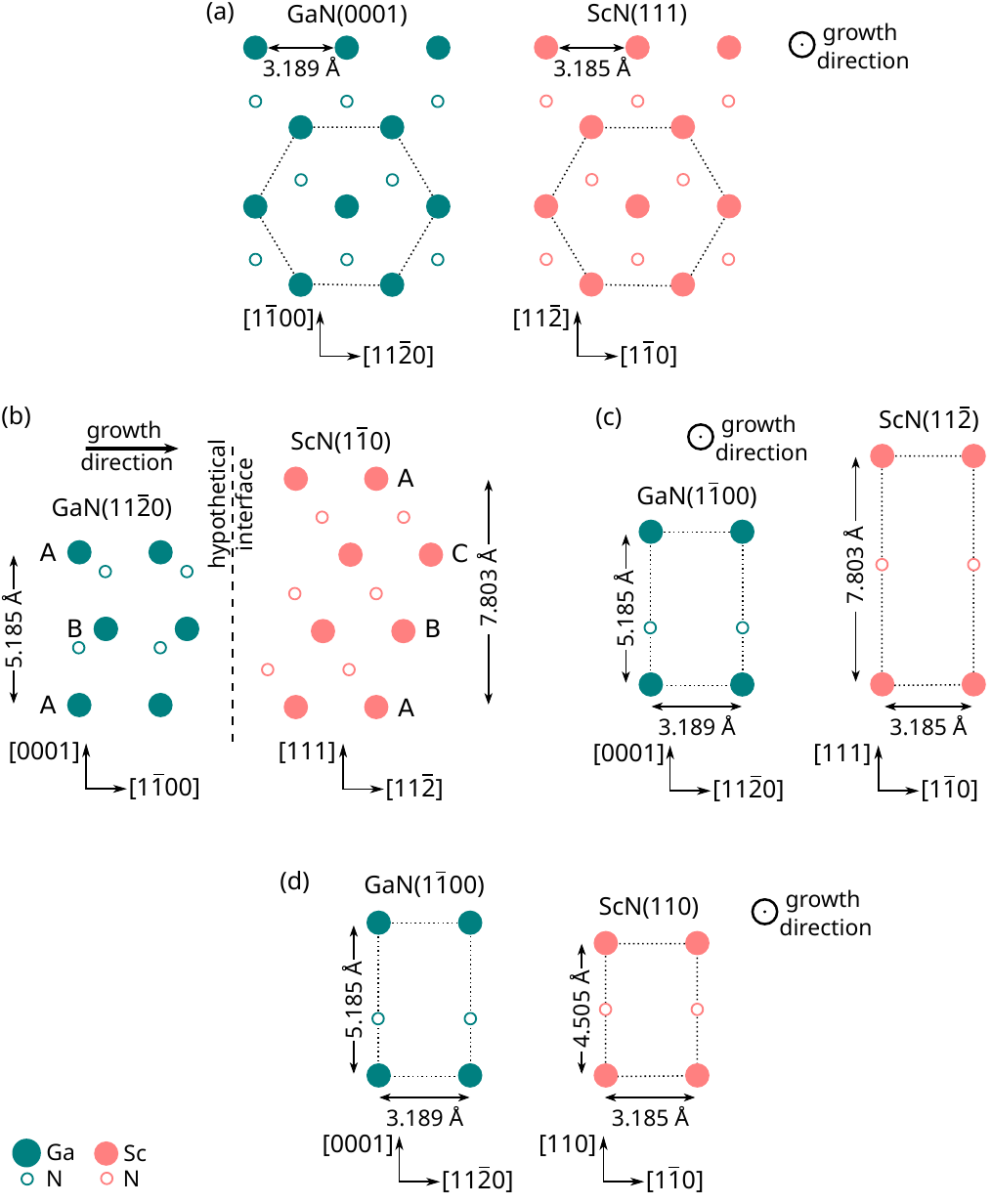}
	\caption{(a) Schematic two-dimensional GaN(0001) and ScN(111) unit meshes. (b) Schematic atomic arrangement of a hypothetical GaN(1\={1}00)/ScN(11\={2}) interface (in cross-sectional view), where (c) shows the corresponding GaN(1\={1}00) and ScN(11\={2}) plan-view unit meshes, which would need to be superimposed to keep the epitaxial relationship that is found for ScN on GaN(0001). (d) Two-dimensional GaN(1\={1}00) and ScN(110) unit meshes, showing the actual epitaxial relationship obtained in this work.}
	\label{Fig1}
\end{figure*}

The common orientation of epitaxial ScN on \emph{C}-plane GaN, as also observed in this work for ScN grown on GaN nanowire top facets, is described by ScN(111)[1\={1}0]$||$GaN(0001)[11\={2}0]. The corresponding GaN(0001) and ScN(111) unit meshes, schematized in plan-view in Figure\,\ref{Fig1}(a), exhibit the same atomic arrangement and quasi identical lattice parameters. The superposition of these planes thus results in a commensurate interface with negligible epitaxial strain. Nevertheless, twinning is inevitable due to the different GaN(0001) and ScN(111) rotational symmetries (6-fold and 3-fold, respectively), which are defined by the stacking of the subsequent monolayers not visible in the unit meshes. 

If the same orientation was kept for ScN grown on the GaN(1\={1}00) surface, the corresponding orientation-relationship would be described by ScN(11\={2})[1\={1}0]$||$GaN(1\={1}00)[11\={2}0], leading to a hypothetical interface as schematized in Figure\,\ref{Fig1}(b) in cross-sectional view. Although the inter-planar lattice distances along the GaN[0001] and ScN[111] directions match, the mismatch in $AB_\text{GaN}$ and $ABC_\text{ScN}$ stacking sequence leads to different lattice parameters. Epitaxy in this orientation would result in different bond lengths and angles between the atoms of both materials, thus preventing a simple atomic configuration at the interface and likely leading to a high density of dangling bonds. Indeed, the dimensions of the corresponding GaN(1\={1}00) and ScN(11\={2}) plan-view unit meshes shown in Figure\,\ref{Fig1}(c) show an enormous mismatch ($>50$\,\%) along the GaN[0001] and ScN[111] directions. The superposition of these planes does not lead to a reasonable interface, making a ScN(11\={2}) orientation on GaN(1\={1}00)  unlikely.

Instead, the epitaxial relationship experimentally observed in this work is given by ScN(110)[001]$||$GaN(1\={1}00)[0001], as explained in the main manuscript and reproduced in Figure\,\ref{Fig1}(d). Although this orientation involves a $13.1$\,\% uniaxial lattice mismatch, the fast strain relaxation via the coincidence site lattice including a periodic array of misfit dislocations (see main manuscript) seems to lead to an energetically more favorable interface.

%======================================================== 
